\documentclass[hyper]{JHEP} 

\usepackage{epsfig}

\newcommand\fverb{\setbox\pippobox=\hbox\bgroup\verb}

\newcommand\fverbdo{\egroup\medskip\noindent%

            \fbox{\unhbox\pippobox}\ }

\newcommand\fverbit{\egroup\item[\fbox{\unhbox\pippobox}]}

\newbox\pippobox

\title{On the Universal Tachyon and
Geometrical Tachyon}

\author{J. Kluso\v{n}
 \footnote{On leave from Masaryk University, Brno}\\
Dipartimento di Fisica \& Sezione I.N.F.N.\\
Universit\`a di Roma
``Tor Vergata'' \\
Via della Ricerca Scientifica 1 00133  Roma   ITALY\\
E-mail:
\email{Josef.Kluson@roma2.infn.it}}
\author{Kamal L. Panigrahi\\
Department of Physics\\
Indian Institute of Technology, Guwahati, 781 039, Guwahati, INDIA \\
E-mail: \email{panigrahi@iitg.ernet.in}} \preprint{{0704.3013
[hep-th]}} \abstract{We study  properties of non-BPS D$(p+1)$-brane
in the background of $k$ NS5-branes, with one transverse direction
compactified on a circle, from the point of view of
Dirac-Born-Infeld action. We present the analysis of two different
embedding of non-BPS D$(p+1)$-brane in given background and study
the classical solutions of given world-volume theory. We argue for
the configuration of a non-BPS D$(p+1)$-brane which allows us to
find solutions of the equations of motion that give unified
descriptions of $G$ and $U$-type branes.}

\def\tr{\tilde{r}}

\def\mT{\mathcal{T}}

\def\tX{\tilde{X}}

\def\mV{\mathcal{V}}

\def\mG{\mathcal{G}}
\def\mGi{\left(\mG^{-1}\right)}
\newcommand{\tR}{\tilde{R}}
\newcommand{\ty}{\tilde{y}}
\newcommand{\tY}{\tilde{Y}}
\begin{document}
\section{Introduction}\label{first}
Non-BPS branes are important and useful objects in string
theory \cite{Sen:1999mg}. The BPS D-branes can be seen as some kind
of soliton solutions on the worldvolume theory of non-BPS branes.
Type IIA/IIB string theories contain odd/even unstable non-BPS
branes in its spectrum. The most important feature of the non-BPS
D-branes is that it contains in its open string spectrum a single
mode of negative ${\rm mass^2}$ particle in addition to an
infinite number of ${\rm mass^2 \ge 0}$ modes. Unlike the BPS
D$p$-brane the non-BPS brane is neutral under the $(p+1)$ form
gauge field. The study of the physics of tachyonic mode has been
one of the most interesting and wide spread subject in the last
few years (for a detailed review and for a comprehensive list of
references see\cite{Sen:2004nf}). In the study of tachyon dynamics
on the D-branes the Dirac-Born-Infeld (DBI) analysis surprisingly
captures well many aspects of the full theory itself
\cite{Sen:1999md,Garousi:2000tr,Bergshoeff:2000dq,Kluson:2000iy,
Kutasov:2003er,Niarchos:2004rw}.

However there has not been any clear cut understanding of the
geometric picture of these open string tachyonic modes. Not too
long back Kutasov \cite{Kutasov:2004dj, Kutasov:2004ct} gave a
geometric interpretation of perturbative open string tachyon in a
D-brane system with a different kind of instability. This involves
a system of coincident $k$ number of NS5-branes and a BPS
D$p$-brane placed at a distance opposite to the NS5-branes. This
configuration is non-supersymmetric as the NS5-brane and D-brane
break different halves of the type II supersymmetry. Many aspects
of the dynamics of this system have been investigated in (see for
example
\cite{Nakayama:2004yx,Yavartanoo:2004wb,Ghodsi:2004wn,Sahakyan:2004cq,
Kluson:2004xc,
Kluson:2004yk,Thomas:2004cd,Chen:2004vw,
Nakayama:2004ge,Thomas:2005fw} and for recent review
\cite{Nakayama:2007sb}.).
Particularly in \cite{Kutasov:2004ct} it was observed that the
dynamics of D$p$-brane in the background of $k$ NS5-branes on a
transverse $R^3 \times S^1$ is remarkably similar to that of the
BPS and non-BPS branes in ten dimensional spacetime. It was found
out that the object which looks like BPS and non-BPS branes to a
six dimensional observer is the same BPS D$p$-brane wrapped or
unwrapped around the extra $S^1$. Various interesting facts about
the non-BPS D-brane in the background of NS 5-brane wrapped along
the transverse $R^3\times S^1$ has been discussed in detail in
\cite{Kluson:2004yk}.

Recently Sen \cite{Sen:2007cz} \footnote{See also
\cite{Israel:2007zc}.} argued that under certain conditions these
BPS D$p$-branes with geometric instability due to its placing at a
saddle point of the potential are identified with a non-BPS
D-brane with the usual open string tachyon instability thereby
giving an interesting geometric meaning. The non-BPS
D$(p+1)$-branes considered in \cite{Sen:2007cz} are the ones where
the worldvolume directions are along the transverse circle (either
wrapping it or ending on NS 5-brane) and along the directions
parallel to the D$p$-branes in the NS5-brane background. Clearly
these D-branes are unstable because of the usual tachyon which
lives in their worldvolume. For a large radius of the transverse
circle $S^1$ the descent relation among various unstable
D$p$-brane system has been discussed in detail. In particular two
types of D-branes, namely the $G$-type D-brane and $U$-type
D-brane seems to be interesting indeed. The G-type of branes have
the `geometrical' instability \cite{Kutasov:2004ct} are obtained
by placing the usual BPS D$p$-brane $(p\le 5)$ along
$x^0,\cdots,x^p$ at $\vec{r}=0, y=\pi R$
 and arbitrary values of
$x^{p+1},\cdots x^5$
\footnote{In the next section we will
present the notational details.}.
 The non-supersymmetric $U$-type D-branes are
obtained by placing the non-BPS D-brane along $x^0,...,x^p, y$
directions (where $y$ is the compactified transverse directions of
the NS 5-brane). They carry the `usual' open string tachyon.
However it has been found out
in \cite{Sen:2007cz} that
 for the large value of the radius of
the transverse circle, both $G$-type and $U$-type D-branes are
actually two phases of the same underlying theory-the worldvolume
theory of $U$-type D-brane.

In the present paper, we would like to investigate the recent
conjecture of Sen from the view point of Dirac-Born-Infeld (DBI)
analysis. We present two complementary descriptions of a non-BPS
D$p$-brane. The first one is a direct DBI analysis of the
configuration studied recently in \cite{Sen:2007cz}. Explicitly,
we present DBI analysis of the non-BPS D$(p+1)$-brane that wraps
the transverse circle. We will argue for the existence of
classical tachyon solutions that reproduce the D-brane
configurations studied recently in \cite{Sen:2007cz}.

The second approach is based on the analysis presented some time
back in \cite{Kluson:2004yk}. We show that non-BPS D$(p+1)$-brane
that is stretched along the world-volume directions of NS5-brane
contains in its solution the non-BPS D$(p+1)$-brane that wraps the
transverse circle and also contains a solution which corresponds
to $G$-type D-brane which sits at the point $Y=\pi R$. We can find
these solutions very easily using the mapping of the mode that
parameterises the position of a non-BPS D$(p+1)$-brane along
$y$-direction to the new ``geometric'' tachyon field $\mT$
\cite{Kutasov:2004dj, Kutasov:2004ct} and then using the profound
analysis pioneered in \cite{Sen:2003tm}. We believe that the
result presented in this paper is the first indication of the
unified description of the $G$ and $U$-branes using the
world-volume theory on non-BPS D$(p+1)$-brane that is localised on
$y$-circle.

This correspondence can be seen more clearly when we perform zero
radius limit analysis following \cite{Kutasov:2004ct, Sen:2007cz}.
Explicitly, we  show that for $k=2$ ($k$ is the number of NS
5-branes) the geometric tachyon $\mT$ and the open string tachyon
$T$ have completely the same dynamics. Since these two type of
D-branes arise as solutions of the same theory they, from the
point of view of this theory, are indistinguishable, which is
identical to Sen's observation. In other words, for $k=2$ the
unstable D$(p+1)$-brane wrapping the transverse circle and the
$G$-type D$p$-brane sitting at the point $\tY=\pi$
can be interpreted as the same object in the zero radius limit.
 This can be thought of as a support of Sen's recent
conjecture\cite{Sen:2007cz}.

Rest of the paper is organised as follows. In section-2, we
present various unstable D-brane configuration in NS 5-brane
background from the view point of Dirac-Born-Infeld analysis. We
show the existence of classical tachyon solutions which correspond
to various unstable D-brane configurations. In section-3, we study
the non-BPS D$(p+1)$-brane extended along the worldvolume
directions of NS 5-branes. We find out a unified description of
the $G$-type and $U$-type D-brane configurations by using the
worldvolume theory on the non-BPS D$(p+1)$-brane in NS5-brane
background that as opposite to Sen's approach is localised at the
point $Y=\pi R$. Section-4 is devoted to the study of the
correspondence in the zero radius limit.  We again find that the
non-BPS D$(p+1)$-brane localised at the point $\tY=\pi$ gives
unified and natural description of $U$ and $G$-unstable D-branes.
We finish the paper in section-5 with conclusions and discussions.

\section{Unstable D-brane Configurations from DBI Point of View
in NS5-brane Background }
Let us consider a system of $k$ NS5-branes in type IIA/IIB theory
stretched along $(x^0,\dots,x^5)$ plane and placed at
$(x^6,\dots,x^9)=(0,\dots,0)$. Let $x^6$ be a compact coordinate
with the periodicity $2\pi R$. The string frame metric, the
dilaton $\Phi$ and the NS sector $3$ form field strength $H$
produced by this background are given by \footnote{We use
$\alpha'=1$ in this paper.}
\begin{eqnarray}\label{ans}
ds^2&=&\eta_{\mu\nu}dx^\mu dx^\nu
+H(\vec{r},y)(dy^2+d\vec{r}^2) \ ,
\nonumber \\
e^{2\Phi}&=&g^2 H(\vec{r},y)\ ,
\nonumber \\
H_{mnp}&=&-\epsilon_{mnpq}\partial^q\Phi  \ ,
\nonumber \\
\end{eqnarray}
where $\mu,\nu$ run from $0$ to $5$; $m,n,p,q$ run from $6$ to $9$
\begin{equation}
\vec{r}=(x^7,x^8,x^9)\ , \quad  y \equiv  x^6
\end{equation}
and
\begin{equation}
H(\vec{r},y)=
1+\frac{k}{2Rr}
\frac{\sinh (r/R)}
{\cosh(r/R)-\cos(y/R)} ,
\quad r\equiv |\vec{r}| \ .
\end{equation}
It is easy to see that the background above is invariant under the
following transformations
\begin{equation}
\sigma: y\rightarrow -y, \quad
\vec{r}\rightarrow -\vec{r}\ .
\end{equation}
If the $k$ 5-branes  are not coincident but are placed at
different points $(\vec{r}_i,y_i)$ $(1\leq i\leq k)$ then the
solution is still described by (\ref{ans}) but with the $H$ given
by
\begin{equation}
H(\vec{r},y)=
1+\sum_{i=1}^k
\frac{1}{2R|\vec{r}-
\vec{r}_i|}\frac{\sinh
(|\vec{r}-\vec{r}_i|/R)}
{\cosh(|\vec{r}-\vec{r}_i|/R)-
\cos ((y-y_i)/R)} \ .
\end{equation}
Let us now consider a non-BPS D$(p+1)$-brane embedded in this
background. Recall that the action for this system is
\cite{Sen:1999md,Garousi:2000tr,Bergshoeff:2000dq,Kluson:2000iy,
Kutasov:2003er,Niarchos:2004rw}
\begin{equation}\label{actn}
S=-\sqrt{2}\mT_{p+1}
\int d^{p+2}\xi e^{-\Phi}
V(T)\sqrt{-\det\mG} \ ,
\end{equation}
where
\begin{equation}
\mG_{\alpha\beta}=
g_{MN}\partial_\alpha X^M\partial_\beta X^N+
b_{MN}\partial_\alpha X^M\partial_\beta X^N
+\partial_\alpha T\partial_\beta T \ ,
\end{equation}
and where now $\left\{\xi^\alpha\right\} (0\leq \alpha \leq p+1)$
are the D$(p+1)$-brane world-volume coordinates and $X^M$
parameterise embedding of D$(p+1)$-brane in target space where
$M,N=0,\dots,9$. Finally $\mT_{p+1}$ is BPS D$(p+1)$-brane
tension.

Note that the equations of motion for $X^M$
 that follow from (\ref{actn})
take the form
\begin{eqnarray}\label{eqXM}
& &\partial_M e^{-\Phi}
V(T)\sqrt{-\det\mG}
+\nonumber \\
&+&\frac{1}{2}e^{-\Phi}V(T)
(\partial_M g_{NK}+
\partial_M b_{NK})\partial_\alpha
X^N\partial_\beta X^K\mGi^{\beta\alpha}
\sqrt{-\det\mG}-
\nonumber \\
&-&\partial_\alpha
[e^{-\Phi}V(T) g_{MN}\partial_\beta X^N
\mGi^{\beta\alpha}_S\sqrt{-\det\mG}]
-\nonumber \\
&-&\partial_\alpha
[e^{-\Phi}V b_{MN}\partial_\beta X^N
\mGi^{\beta\alpha}_A\sqrt{-\det\mG}]=0 \ ,
\nonumber \\
\end{eqnarray}
where
\begin{equation}
\mGi^{\alpha\beta}_S=
\frac{1}{2}\left(\mGi^{\alpha\beta}+
\mGi^{\beta\alpha}\right)\,
\quad
\mGi^{\alpha\beta}_A=
\frac{1}{2}\left(\mGi^{\alpha\beta}-
\mGi^{\beta\alpha}\right) \ .
\end{equation}
On the other hand the equation of motion
for tachyon is given by
\begin{equation}\label{eqT}
\frac{dV(T)}{dT}
e^{-\Phi}
\sqrt{-\det\mG}
-\partial_\alpha [e^{-\Phi}
V(T)\partial_\beta T\mGi^{\beta\alpha}_S
\sqrt{-\det\mG}]=0 \ .
\end{equation}
Let us now  consider  following ansatz
\begin{equation}\label{NBPSwa}
X^i=\xi^i \ , i=0,\dots,p,  \quad
Y=\xi^{p+1}\ , \quad T=T(y) \
\end{equation}
that corresponds to non-BPS D$(p+1)$-brane spanning the
coordinates $x^0,\dots,x^p$ and $y$-direction. Note that for this
ansatz the components of the matrix $\mG$ take the form
\begin{eqnarray}
\mG_{ij}&=&
\eta_{ij}+H(\vec{R},y)\partial_i
\vec{R}\partial_j \vec{R} \ , \nonumber \\
\mG_{yy}&=&H(\vec{R},y)+
H(\vec{R},y)\partial_y \vec{R}
\partial_y \vec{R}+\partial_y T
\partial_y T  \ ,  \nonumber \\
\mG_{yi}&=&\mG_{i y}=
H(\vec{R},y)\partial_i \vec{R}
\partial_y \vec{R} \ , \nonumber \\
\end{eqnarray}
where we have not specified the form of the world-volume
dependence of the modes $\vec{R}$.

It is easy to see that when we
insert (\ref{NBPSwa}) into the
equations of motions for $X^i$  (\ref{eqXM}),
these equations are
satisfied automatically.
On the other hand, the equation of motion
(\ref{eqXM}) for $X^M\equiv Y$ implies
\begin{eqnarray}\label{eqXMY}
& &\partial_y e^{-\Phi}
V(T)\sqrt{-\det\mG}
+\frac{1}{2}e^{-\Phi}
V(T)\partial_y g_{yy}\mGi^{yy}
\sqrt{-\det\mG}-
\nonumber \\
&-&\partial_y
[e^{-\Phi}V(T) g_{yy}
\mGi^{yy}_S\sqrt{-\det\mG}]=0 \ .
 \nonumber \\
\end{eqnarray}
Further, the equations of motion for $\vec{R}$
imply
\begin{eqnarray}\label{eqXm}
& &\partial_{X^m} e^{-\Phi}
V(T)\sqrt{-\det\mG}+\nonumber \\
&+&\frac{1}{2}e^{-\Phi}V(T)
(\partial_{x^m} g_{NK}+
\partial_{x^m} b_{NK})\partial_\alpha
X^N\partial_\beta X^K\mGi^{\beta\alpha}
\sqrt{-\det\mG}-
\nonumber \\
&-&\partial_\alpha
[e^{-\Phi}V(T) g_{mn}\partial_\beta X^n
\mGi^{\beta\alpha}_S\sqrt{-\det\mG}]-
\nonumber \\
&-&\partial_\alpha
[e^{-\Phi}V(T) b_{mn}\partial_\beta X^n
\mGi^{\beta\alpha}_A\sqrt{-\det\mG}]=0\ ,
\nonumber \\
\end{eqnarray}
where $ m=7,8,9$.
To simplify the discussion further let us now presume that
$\vec{R}$ is constant. Then
\begin{eqnarray}
\mG_{ij}=\eta_{ij} \ , \quad
\mG_{yy}=H(\vec{R},y)+(\partial_yT)^2 \ , \quad
\mG_{yi}=0 \ . \nonumber \\
\end{eqnarray}
For this ansatz the equations of motion for $X^m$ (\ref{eqXm})
simplifies as
\begin{eqnarray}\label{eqXma}
\frac{\partial_{x^m}H}
{ H^{3/2}}\frac{(\partial_yT)^2}
{\sqrt{H+(\partial_yT)^2}}=0 \ .
\nonumber \\
\end{eqnarray}
We see that  for $\partial_y T\neq 0$ the only
possible solution of the
equation above occurs
for $\partial_{x^m}H=0$. It turns out that
there is no solutions of
this equation for general $\vec{r}_i$. In
what follows we restrict
ourselves to the case of $\vec{r}_i=0$.
Then it is easy to see that the solution of the equation
$\partial_{x^m}H=0$ is $\vec{R}=0$.

On the other hand in
case of the constant tachyon $T={\rm{const}}$,
we obtain that the equation  (\ref{eqXma}) is solved for any
$\vec{R}=\vec{c}$ and for general $\vec{r}_i$. This is
exactly the same
solution that was  found recently
 in \cite{Sen:2007cz}.

Let us now consider
the equation (\ref{eqXMY}) in more detail. For
(\ref{NBPSwa}) and for $\vec{R}=0$ it implies
\begin{eqnarray}\label{eqXMYr}
-\frac{\partial_y H(y)V(T)}{2 H^{3/2}(y)}
\frac{(\partial_yT)^2}{\sqrt{H(y)
+(\partial_yT)^2}}
-\partial_y
[\frac{V(T) H^{1/2}(y)}{\sqrt{H(y)+
(\partial_yT)^2}}]=0 \ ,
 \nonumber \\
\end{eqnarray}
where
\begin{eqnarray}
H(y)\equiv\lim_{r\rightarrow 0}H=
1+\frac{k}{2R^2}
\frac{1}{1-\cos y/R} \ .  \nonumber \\
\end{eqnarray}
Finally, the equation of
motion for $T$ (\ref{eqT}) reduces, for
$\vec{R}=0$, into
\begin{eqnarray}\label{eqTr}
\frac{dV(T)}{dT}
\frac{H^{1/2}(y)}
{\sqrt{H(y)+(\partial_yT)^2}}
-V(T)\partial_y
\left[\frac{ \partial_yT}{H^{1/2}(y)
\sqrt{H(y)+(\partial_yT)^2}}\right]=0 \ .
\nonumber \\
\end{eqnarray}
Clearly all these equations have the solution where
$\frac{dV}{dT}=0$ and $T=\mathrm{const}$. For $T_{max}=0$ that
corresponds to $V(T_{max})=1$  we obtain an unstable
D$(p+1)$-brane that wraps $y$-circle and that is generally
localised at an arbitrary $\vec{R}=\vec{c}$ as we have argued
above. On the other hand for the minimum of the potential that
occurs at  $T_{min}=\pm \infty, \quad
V(T_{min})=0$ this configuration corresponds to
the closed string vacuum after the tachyon condensation.

Our goal is to consider more general solutions. We will argue that
in case of the non-BPS D$(p+1)$-brane wrapping $y$ direction there
exists solution with natural physical meaning that is however
 singular.
 Let us introduce the `Heaviside step
function' $\mathcal{H}(x)$ defined as
\begin{equation}
\mathcal{H}(x)=\left\{\begin{array}{ccc}
0 \quad \mathrm{for} \ x<0 \\
\frac{1}{2} \quad \mathrm{for} \ x=0 \\
1 \quad \mathrm{for} \ x>0 \\
\end{array}\right. \ .
\end{equation}
The characteristic property of this function
is
\begin{equation}
\frac{d}{dx}\mathcal{H}(x)=\delta(x) \ .
\end{equation}
Now let us consider following form of the
tachyon
\begin{equation}\label{tachsols}
T(y)=-|T_{min}|+2|T_{min}|\mathcal{H}(y-\pi R) \ .
\end{equation}
The physical
meaning of this ansatz is clear. For $0<y<\pi
R$ we have $T=-T_{min}$;
for $y=\pi R$ we have $T=0$ and for $\pi
R<y<2\pi R$ we have $T=T_{min}$.
It is clear that this tachyon
condensation corresponds to D$p$-brane that
is localised at $y=\pi
R$ and to anti D$p$-brane that is localised at $y=0$. More
precisely, for this solution we have
\begin{equation}
\frac{dV}{dT}=0
\end{equation}
for all $y$.
Explicitly, let us again write equation of motion for
the tachyon (\ref{eqTr})
\begin{eqnarray}
\frac{dV}{dT}
\frac{H^{1/2}}
{\sqrt{H+(\partial_yT)^2}}
-V\partial_y
\left[\frac{ \partial_yT}{H^{1/2}
\sqrt{H+(\partial_yT)^2}}\right]=0 \ .
\nonumber \\
\end{eqnarray}
Firstly, it is clear that the ansatz
(\ref{tachsols})
 solves the equation of motion
for $y\neq 0,\pi R$.
For $y=\pi R$, $H(\pi R)$ is finite and
can be neglected with respect to $\partial_yT$.
Then the term proportional
to $\frac{dV}{dT}$ is zero
and the second one is equal to
\begin{equation}
V(T=0)\partial_y\left[\frac{\partial_yT}
{H^{1/2}\sqrt{H+ (\partial_yT)^2}}\right]
\approx
\frac{\partial_y H}{H^{3/2}}=
0 \ .
\end{equation}
In the same way we can argue that the
ansatz (\ref{tachsols}) solves the
equations of motion for $y=0$ as well.
If we insert this ansatz into
the non-BPS D$(p+1)$-brane action we
obtain
\begin{eqnarray}
S(T(y))=
-\sqrt{2}\mT_{p+1}\int_0^{2\pi R}d^{p+1}\xi
dy \frac{1}{\sqrt{H(y)}}
 V(T)\sqrt{H(y)+(\partial_yT)^2}= \nonumber \\
=-\sqrt{2}\mT_{p+1}|T_{min}|\int d^{p+1}\xi
\left(\frac{1}{\sqrt{H(\pi R)}}
+\frac{1}{\sqrt{H(0)}}
\right) \ .
\nonumber \\
\end{eqnarray}
The result given above suggests that
 the tachyon solution (\ref{tachsols})
describes the configuration of D$p$-brane localised at $y=\pi R$
and anti-D$p$-brane localised at $y=0$ \footnote{More precisely,
in order to distinguish D$p$-brane from anti-D$p$-brane we should
consider the Wess-Zumino term for a non-BPS D$p$-brane
\cite{Okuyama:2003wm,Takayanagi:2000rz,Kraus:2000nj} that is
proportional to $\int dT\wedge C$ where $C$ is a collection of
Ramond-Ramond forms. The presence of $dT$ determines whether the
tachyon kink is D$p$-brane or anti D$p$-brane.}. On the other hand
we see that the solution is singular due to the fact that it is
defined using the  function $\mathcal{H}$.
 Further, due
to the presence of $|T_{min}|$ in the final result it is not
completely clear that the solution above really describes BPS
D$p$-branes.  On the other hand we present in next section
 a more natural solution of this problem.

Let us now consider
 another possibility of the tachyon condensation on
non-BPS D$(p+1)$-brane wrapping $y$-circle and consider  following
ansatz
\begin{equation}\label{anstT2}
X^i=\xi^i , \quad
i=0,\dots,p-1 \ , \quad
X^p=\xi^p \equiv x \ ,  \quad
Y=\xi^{p+1} \ , \quad
T=f(x) \ ,
\end{equation}
where $f(u)$ is a function with the properties
\begin{equation}\label{anstT2f}
f'(u)>0 \ , f(\pm \infty)=
\pm \infty \ ,
\end{equation}
and where $a$ is a constant that is taken to
infinity in the end.

For the ansatz (\ref{anstT2})
the components of the matrix $\mG$
takes the form
\begin{eqnarray}
\mG_{ij}&=&
\eta_{ij}+H(\vec{R},y)\partial_i
\vec{R}\partial_j \vec{R} \ , \nonumber \\
\mG_{yy}&=&H(\vec{R},y)+
H(\vec{R},y)\partial_y \vec{R}
\partial_y \vec{R} \ ,  \nonumber \\
\mG_{yi}&=&\mG_{i y}=
H(\vec{R},y)\partial_i \vec{R}
\partial_y \vec{R} \ , \nonumber \\
\mG_{pp}&=&1+H(\vec{R},y)
\partial_x \vec{R}\partial_x\vec{R}
+ (\partial_x T)^2 \ , \nonumber \\
\mG_{pi}&=&\mG_{i p}=
H(y,\vec{R})\partial_x\vec{R}\partial_i
\vec{R} \ , \nonumber \\
\mG_{yp}&=&\mG_{py}=
H(\vec{R},y)\partial_y\vec{R}\partial_x \vec{R} \ .
\nonumber \\
\end{eqnarray}
We again solve the equations of motion
for constant $\vec{R}$. For this ansatz
the matrix $\mG$ have following components
\begin{eqnarray}
\mG_{ij}&=&
\eta_{ij} \ , \quad
\mG_{yy}=H(\vec{R},y) \ , \quad
\mG_{yi}=\mG_{i y}=0 \ , \nonumber \\
\mG_{pp}&=&1
+ (\partial_x T)^2 \ , \quad
\mG_{p i}=\mG_{i p}=
\mG_{yp}=\mG_{py}=0 \ ,
\nonumber \\
\det\mG&=&-H(\vec{R},y)
(1+(\partial_x T)^2) \ .
\nonumber \\
\end{eqnarray}
Then the equations of motion
(\ref{eqXM})  for $
X^m$ take the form
\begin{eqnarray}
\frac{\partial_{x^m}H}
{H\sqrt{1+(\partial_x T)^2}}(\partial_xT)^2=0
\nonumber \\
\end{eqnarray}
that implies that the only
solution for $\partial_x T\neq 0$ is
$\partial_{x^m}H=0$ which occurs for $X^m=0$.
The equation of
motion for $y$ takes the form
\begin{eqnarray}\label{eqyp}
&-&\frac{1}{2g H^{3/2}}
\partial_y HV(T)\sqrt{H(1+
(\partial_x T)^2)}
+\frac{V(T)\partial_y H}{2g H^{1/2}
\sqrt{H(1+ (\partial_x T)^2)}}
-\nonumber \\
&-&\partial_y
[\frac{V(T) H}{2g H^{1/2}H}\sqrt{H(1+
(\partial_x T)^2)}]
-\partial_x
[\frac{V(T) H}{2g H^{1/2}\sqrt{H(1+
(\partial_x T)^2)}}]=
\nonumber \\
&=&-\frac{1}{2g}\frac{
V(T)\partial_y H (\partial_x T)^2}
{H\sqrt{1+(\partial_x T)^2}}
-\frac{1}{2g}
\partial_x\left[\frac{V(T)}{\sqrt{1+
(\partial_x T)^2}}\right]=0 \ .
\nonumber \\
\end{eqnarray}
Further, the equation of motion
for tachyon implies
\begin{eqnarray}
& &\frac{dV(T)}{dT}
\frac{1}{g\sqrt{H}}
\sqrt{H(1+(\partial_x T)^2)}
-\partial_x
[\frac{V(T)\partial_x T\sqrt{H}}
{g\sqrt{H}\sqrt{1+(\partial_x T)^2}}
]=0\nonumber \\
\nonumber \\
\end{eqnarray}
and consequently
\begin{equation}
\partial_x
\left[\frac{V(T)}{\sqrt{1+(\partial_x T)^2}}
\right]=0 \
\end{equation}
that implies
\begin{equation}\label{TC}
\frac{V(T)}{\sqrt{1+(\partial_xT)^2}}=C \ .
\end{equation}
Following \cite{Sen:2003tm}
we can argue that the constant $C$ has
to vanish. In fact, for kink
solution $T\rightarrow \pm \infty$
for $x\rightarrow \pm \infty$ where however
$\partial_x T$ is finite but
$\lim_{T\rightarrow \pm \infty}V=0$.
Then we get that $C=0$ and clearly
the tachyon given in (\ref{anstT2}) and
in (\ref{anstT2f}) implies $C=0$ in the
limit $a\rightarrow \infty$.
Further, we now argue that the tachyon
profile given in (\ref{anstT2}) solves
the equation of motion
(\ref{eqyp}). Firstly, it is clear
that the second term in (\ref{eqyp}) vanishes
for (\ref{TC}). On the other hand
the first term for $a\rightarrow
\infty$ is proportional to
\begin{equation}
-\frac{1}{2g}\frac{
V(T)\partial_y H (\partial_x T)^2}
{H\sqrt{1+(\partial_x T)^2}}
\approx
\frac{V(f(ax)) a^2f'^2(ax)}
{|af'(ax)|}\approx
\lim_{a\rightarrow \infty}
 e^{-\frac{f(ax)}{\sqrt{2}}
}a \rightarrow 0
\end{equation}
in the limit $a\rightarrow \infty$ since
we presume that the tachyon potential
behaves for large $T$ as $e^{-\frac{T}{\sqrt{2}}}$.
 Then
inserting the ansatz (\ref{anstT2}) into the non-BPS
D$(p+1)$-brane action we obtain
\begin{eqnarray}\label{Sanst2}
S&=&-\frac{\sqrt{2}\mT_{p+1}}{g}
\int d^p\xi dy dx
 V(T(ax))
\sqrt{1+(\partial_x T)^2}\approx\nonumber \\
&\approx&
-\frac{\sqrt{2}\mT_{p+1}}{g}\int dx V(f(ax))
af'(ax)\int d^p\xi dy=\nonumber \\
&=&
-\frac{\sqrt{2}\mT_{p+1}}{g}
\int dm V(m) 2\pi R
\int d^p\xi
=-\frac{2\pi R\mT_p}{g}\int d^p\xi
\ ,
\nonumber \\
\end{eqnarray}
where in the second step we have taken
the large $a$ limit and in the last step
we have used the fact that
\cite{Sen:2003tm}
\begin{equation}\label{tachV}
\sqrt{2}\mT_{p+1}\int dm V(m)=
\mT_p \ .
\end{equation}
 (\ref{Sanst2}) suggests
following physical interpretation of the
ansatz (\ref{anstT2}): The tachyon kink solution (\ref{anstT2})
can be interpreted as  a  D$p$-brane which is
localised at point $x^p=0$
and is extended in $x^0,\dots,x^{p-1},y$.
This is precisely the
solution presented in \cite{Kutasov:2004ct}.
\section{Non-BPS D$(p+1)$-brane Extended
Along Worldvolume
of $k$ NS5-branes}
The main goal of our paper is to give an unified
description of the configurations presented in
\cite{Kutasov:2004ct,Sen:2007cz}.
We give a description that is based
on the analysis given in \cite{Kluson:2004yk}.
Explicitly, let us
consider a non-BPS D$(p+1)$-brane
with their world-volume lying
completely along the directions of the NS5-branes. This
configuration is  achieved with
the following ansatz
\begin{equation}
X^\alpha=\xi^\alpha \ , \quad  \alpha=0,\dots,p+1 \
\end{equation}
and hence
 the action for a non-BPS D$(p+1)$-brane takes
the form
\begin{eqnarray}\label{SnonBPS}
S&=&-\sqrt{2}\mT_{p+1}
\int d^{p+2}\xi
\frac{1}{\sqrt{H(\vec{R},Y)}}
V(T)\sqrt{-\det\mG} \ ,
\nonumber \\
\mG_{\alpha\beta}&=&
\eta_{\alpha\beta}+H(\vec{R},Y)
\partial_\alpha \vec{R}
\partial_\beta \vec{R}+
H(\vec{R},Y)\partial_\alpha Y
\partial_\beta Y+\partial_\alpha T
\partial_\beta T \ .
\nonumber \\
\end{eqnarray}
Again solving the equation of motion
for constant $X^m$ gives the solution
(for coincident NS5-branes)
\begin{equation}
\vec{R}=0 \ .
\end{equation}
Using this result the
 action (\ref{SnonBPS}) simplifies
 considerably
\begin{eqnarray}\label{actYT}
S&=&-\sqrt{2}\mT_{p+1}
\int d^{p+2}\xi
\frac{1}{\sqrt{H(Y)}}
V(T)\sqrt{-\det\mG} \ ,
\nonumber \\
\mG_{\alpha\beta}&=&
\eta_{\alpha\beta}+
H(Y)\partial_\alpha Y
\partial_\beta Y+\partial_\alpha T
\partial_\beta T \ ,
\nonumber \\
\end{eqnarray}
where
\begin{equation}
H(y)=\lim_{r\rightarrow 0}H(r,y)=
1+\frac{k}{4R^2\sin^2\frac{Y}{2R}} \ .
\end{equation}
Let us now introduce the geometric
tachyon field $\mT$ through the relation
\begin{equation}\label{mTY}
d\mT=\sqrt{H}dY=\sqrt{1+\frac{k}{4R^2\sin^2
\frac{Y}{2R}}}dY \ .
\end{equation}
Even if it is possible to explicitly
integrate out this equation
the result is not well illuminating.
On the other hand
 we can in principle express
$Y$ as function of $\mT$. Consequently
we can introduce the function $W(\mT)$
defined as
\begin{equation}\label{Wdef}
W(\mT)=\frac{1}{\sqrt{H(\mT)}} \ .
\end{equation}
We will now discuss  basic
properties of the potential $W(\mT)$
defined above. Note that for $Y\rightarrow 0$ we get
$\mT\rightarrow \infty$
and consequently from (\ref{Wdef}) we obtain
$\lim_{\mT\rightarrow \infty}
W(\mT)\rightarrow 0$.
On the other hand for $Y\rightarrow 2\pi
R$ we get $\mT\rightarrow -\infty$
and (\ref{Wdef}) again implies
$\lim_{\mT\rightarrow -\infty}
W(\mT)\rightarrow 0$. Finally
 for $Y\rightarrow \pi R$ we obtain
\begin{equation}
\mT\sim \sqrt{1+\frac{k}{4R^2}}(\pi R-Y)
\end{equation}
and hence for  $\mT\rightarrow 0 (Y\rightarrow
 \pi R)$ we obtain
\begin{equation}
W(\mT=0)=
\frac{1}{g\sqrt{1+\frac{k}{4R^2}}} \ .
\end{equation}
Then the action
(\ref{actYT})-after performing the
field redefinition (\ref{mTY})-
takes the form
\begin{eqnarray}\label{atctW}
S&=&-\sqrt{2}\mT_{p+1}
\int d^{p+2}\xi
W(\mT)
V(T)\sqrt{-\det\mG} \ ,
\nonumber \\
\mG_{\alpha\beta}&=&
\eta_{\alpha\beta}+\partial_\alpha
\mT\partial_\beta \mT+
\partial_\alpha T
\partial_\beta T \ .
\nonumber \\
\end{eqnarray}
 It will be also
useful to  calculate the components
of the world-volume
stress energy tensor
from the action (\ref{atctW}).
In order to do this we
 proceed in the standard way.
Explicitly, we  replace the flat
 worldvolume metric
$\eta_{\alpha\beta}$ with the
general  metric $g_{\alpha\beta}$ and
 define the stress energy tensor as
 \begin{equation}\label{Tcom}
 T_{\alpha\beta}=-
\frac{2}{\sqrt{-g}}\frac{\delta S}
{\delta g^{\alpha\beta}} \ .
\end{equation}
Then from (\ref{atctW}) we obtain that
$T_{\alpha\beta}$ are  equal to
\begin{equation}
T_{\alpha\beta}=-\sqrt{2}\mT_{p+1}
V(T)W(\mT)\eta_{\alpha\gamma}
\eta_{\beta\delta}\mGi^{\delta\gamma}
\sqrt{-\det \mG} \
\end{equation}
using
\begin{equation}
\left.\frac{\delta g_{\gamma\delta}}
{\delta g^{\alpha\beta}}\right|_{g_{\alpha\beta}=
\eta_{\alpha\beta}}=-\eta_{\gamma\alpha}
\eta_{\beta\delta} \ .
\end{equation}
Now using also the fact that the
 world-volume theory defined by
the action (\ref{atctW}) is
Poincare invariant we obtain that the
components of the stress energy
tensor obey the conservation laws
\begin{equation}\label{conT}
\partial_\alpha\eta^{\alpha
\gamma}T_{\gamma\beta}=0 \ .
\end{equation}
Note also that this is an important
difference with respect to the situation
studied in the previous section where
due to the explicit dependence of the
world-volume theory on $y$ the components
of the stress
energy tensor  do not obey
the conservation equation
(\ref{conT}).

Let us now consider the solution
where the tachyon $T$ is in its
unstable minimum $T=0$ while $\mT$
depends on one spatial
coordinate, say $\xi^{p+1}=x$.
Then (\ref{conT}) implies
\begin{equation}
\partial_x T_{xx}=0 \ .
\end{equation}
Since for
$x\rightarrow \pm \infty$ we presume
 $\mT\rightarrow \pm \infty$
and consequently $W\rightarrow 0$ we obtain
that $T_{xx}$ has to
vanish everywhere.
This is well known result given in
\cite{Sen:2003tm}.
Explicitly, following
\cite{Sen:2003tm}
we can consider $\mT$ in the form
\begin{equation}
\mT=f(a x)  \ ,
\end{equation}
where $f(u)$ is a function with the properties
\begin{equation}\label{fprop}
f'(u)>0 \ , f(\pm \infty)=
\pm \infty \ ,
\end{equation}
and where $a$ is a constant that is taken to
infinity in the end. Then it is easy to see that
$T_{xx}$ vanishes for all $x$ in the limit
$a\rightarrow \infty$ since
\begin{equation}
T_{xx}=-\sqrt{2}\mT_{p+1}
\frac{W(\mT)}{\sqrt{1+
(\partial_x\mT)^2}}\approx
-\sqrt{2}\mT_{p+1}
\frac{W(f(ax))}{a|f'(ax)|} \ .
\end{equation}
It is clear that this expression vanishes in the limit
$a\rightarrow \infty$ where $W\rightarrow 0$ and $a\rightarrow
\infty$. This expression also vanishes for $x=0$ since now $W(0)$
is finite but $a\rightarrow \infty$.
Let us now calculate the components of the
kink stress energy tensor. Again, following
\cite{Sen:2003tm} we define these components
as
 \begin{eqnarray}
T_{ij}^{kink}&=&\int dx T_{ij}=
-\eta_{ij}
\sqrt{2}\mT_{p+1}
\int  dx W(f(ax)) a|f'(ax)|=
\nonumber \\
&=&
-\eta_{ij}
\sqrt{2}\mT_{p+1}
\int_{\infty}^{-\infty}
 dm W(m)=
-\eta_{ij}
\sqrt{2}\mT_{p+1}\int_0^{2\pi R}
dy=\nonumber \\
&=&-\eta_{ij}
\sqrt{2}\mT_{p+1}2\pi R=
-\eta_{ij}\sqrt{2}\mT_p R \ ,
\nonumber \\
\end{eqnarray}
where $i,j=0,\dots,p$. In other words the tachyon kink given above
corresponds to non-BPS D$p$-brane that is sitting for $(x<0)$ on
the world-volume of $k$-NS5-branes, and then wraps the circle $y$
of radius $2\pi R$ at $x=0$ and then is again sitting for $x>0$.
We have seen that the tension of the configuration corresponds to
non-BPS D$(p+1)$-brane wrapped $y$-circle and sitting
at $\vec{R}=0$ in the same way as in
paper \cite{Sen:2007cz}.

We can also consider another solution given
by the ansatz
\begin{equation}\label{mtco}
\mT=\mathrm{const}. \ , \quad T=T(x) \ .
\end{equation}
Again  the equation of motion for $\mT$ implies that $\mT=0$ or
$\mT=\pm\infty$. We will consider the ansatz $\mT=0$ corresponding
to non-BPS D$(p+1)$-brane localised at the point $y=\pi R$.  Then
we will construct the
 tachyon singular solution exactly as in
\cite{Sen:2003tm}. Explicitly, let
us consider the tachyon profile
\begin{equation}
 T(x)=f(ax)
 \ ,
\end{equation}
where $f$ has the same properties as
in (\ref{fprop}).
The invariance of the world-sheet
energy tensor implies that
\begin{eqnarray}
\partial_x T_{xx}&=&0 \ , \quad
\quad T_{xx}=-\sqrt{2}\mT_{p+1}W(0)V(T)
\mGi^{xx}\sqrt{-\det\mG}=\nonumber \\
&=&-\sqrt{2}\mT_{p+1}W(0)
V(f(ax))
\frac{1}{\sqrt{1+a^2f'^2(ax)}}
\approx \nonumber \\
&\approx &
-\sqrt{2}\mT_{p+1}W(0)
\frac{V(f(ax))}{a|f'(ax)|}
 \ .
\nonumber \\
\end{eqnarray}
Again standard arguments imply that the
$T_{xx}$ has
to vanish.
On the other hand the remaining components
of the stress energy tensor $T_{ij}$
are equal to
\begin{eqnarray}
T_{ij}&=&-\eta_{ij}\sqrt{2}
\mT_{p+1}W(0)
V(f(ax))
\sqrt{1+a^2f'^2(ax)
}\approx
\nonumber \\
&\approx &-\eta_{ij}\sqrt{2}\mT_{p+1}
\frac{1}{g\sqrt{1+\frac{k}{4R^2}}}
V(f(ax))
a|f'(ax)| \ .
\nonumber \\
\end{eqnarray}
We again define
 the stress energy tensor
of the kink as
\begin{eqnarray}
T_{ij}^{kink}&=&\int dx  T_{ij}(x)=
-\eta_{ij}\sqrt{2}\mT_{p+1}
\frac{1}{g\sqrt{1+\frac{k}{4R^2}}}
\int dx
V(f(ax))a|f'(ax)|=
\nonumber \\
&=&-\eta_{ij}\sqrt{2}\mT_{p+1}
\frac{1}{g\sqrt{1+\frac{k}{4R^2}}}
 \int dm V(m)=
-\eta_{ij}\frac{\mT_p}
{g \sqrt{1+\frac{k}{4R^2}}}
\nonumber \\
\end{eqnarray}
using (\ref{tachV}).
The physical interpretation of the solution
(\ref{mtco})  is as follows. It
 corresponds to BPS D$p$-brane that
is localised at the point $y=\pi R, \vec{R}=0$.
 Again, the tension of this
geometrically unstable D$p$-brane agrees exactly with the results
given in \cite{Kutasov:2004ct,Sen:2007cz}.
\section{Small Radius Limit}
Following \cite{Kutasov:2004ct,Sen:2007cz}
we now consider the limit
\begin{equation}
R\rightarrow 0, \quad  g\rightarrow 0 , \quad
\tilde{g}\equiv \frac{g}{R}=\mathrm{fixed}
\end{equation}
and define new coordinate
\begin{equation}
\tilde{y}=\frac{y}{R},\quad
\vec{\tilde{r}}=\frac{\vec{r}}{R} \ .
\end{equation}
In this limit the NS5-brane background
takes the form
\begin{equation}\label{geR}
ds^2=\eta_{\mu\nu}
dx^\mu dx^\nu+
h(\vec{\tr},\ty)
(d\ty^2+d\vec{\tr}^2)\ ,\quad
e^{2\Phi}=\tilde{g}^2
h(\vec{\tr},\ty) \ ,
\end{equation}
where
\begin{equation}
h(\vec{\tr},\ty)=
\frac{k}{2\tr}\frac{\sinh\tr}
{\cosh\tr-\cos\ty} \ ,
\quad \tr\equiv |\vec{\tr}| \ .
\end{equation}
In this coordinate system the BPS D$p$-brane in unstable
equilibrium is situated at $\vec{\tr}=0,\ty=\pi$. The formulae for
the tension and the tachyon mass of this G-type brane takes the
form \cite{Kutasov:2004ct,Sen:2007cz}
\begin{equation}
\tau_p= \frac{2}{\sqrt{k}} \tilde{g}^{-1}\mT_p \ , \quad
m_T^2=-\frac{1}{k} \ .
\end{equation}
Let us now discuss an unstable D$(p+1)$-brane in the geometry
(\ref{geR}) where we begin with non-BPS D$(p+1)$-brane wrapped
$\ty$ direction. We  restrict to the case of the constant
$\tX^m=0$ that is necessary in case when $\partial_{\ty}T\neq 0$.
Let us consider the equation of motion for $\tY=\xi^{p+1}$
\begin{eqnarray}
-\frac{\partial_{\ty}h}{
h^{3/2}}V(T)\frac{ \partial_{\ty}
T}
{\sqrt{h+(\partial_{\ty}T)^2}}-
\partial_{\ty}\left[\frac{V(T) h^{1/2}}
{\sqrt{h+(\partial_{\ty}T)^2}}\right]=0 \
\nonumber \\
\end{eqnarray}
and consider
following ansatz for tachyon
\begin{equation}\label{Tanssmall}
T(\ty)=-|T_{min}|+2|T_{min}|\mathcal{H}(\ty-\pi) \ .
\end{equation}
Following arguments given in the second section
we can easily shown that $T(\ty)$ solves the
equation of motions.
Inserting (\ref{Tanssmall}) into
a non-BPS D(p+1)-brane action we get
\begin{eqnarray}
S&=&-\frac{\sqrt{2}\mathcal{T}_{p+1}}{\tilde{g}}
\int d^{p+1}\xi d\ty
V(T)\frac{1}{
\sqrt{h}}\sqrt{h+(\partial_{\ty}T)^2}=\nonumber \\
&=&-\frac{\sqrt{2}\mathcal{T}_{p+1}}
{\tilde{g}}|T_{min}|
\int d^{p+1}\xi
\left(\frac{1}{\sqrt{h(\pi)}}
+\frac{1}{
\sqrt{h(0)}}
\right) \ .
\nonumber \\
\end{eqnarray}
In the same way
 as in the second section, it is tempting
to interpret the resulting configuration as a
a BPS D$p$-brane anti-D$p$-brane
localised at $\ty=\pi$ and $\ty=0$ respectively.
However as in the second section we see
that this interpretation is not well
precise due to the
 presence of the
factor $|T_{min}|=\infty$.
In fact,  in order
to derive more natural solution we
proceed in the same way  as in the previous
section.
Let us again  consider
 the non-BPS D$(p+1)$-brane that is embedded
in  $R^{5,1}$ in the geometry
 (\ref{geR}). This can be achieved
with the choice of the gauge
\begin{equation}
X^\alpha=\xi^\alpha \ , \quad   \alpha=0,\dots,p+1 \ .
\end{equation}
In this case the non-BPS D$(p+1)$-brane action takes the form
\begin{eqnarray}\label{nonact2}
S&=&-\frac{\sqrt{2}\mT_{p+1}}{\tilde{g}}
\int d^{p+2}\xi \frac{V(T)}{\sqrt{h(\tR,\tY)}}
\sqrt{-\det \mG} \ , \nonumber \\
\end{eqnarray}
where
\begin{equation}
\mG_{\alpha\beta}=\eta_{\alpha\beta}+
h(\tR,\tY)\left(\partial_{\alpha}\vec{\tR}
\partial_{\beta}\vec{\tR}+
\partial_{\alpha}\tY\partial_{\beta}\tY
\right)
+\partial_{\alpha}T\partial_{\beta}T  \ .
\end{equation}
The action  (\ref{nonact2}) describes the non-BPS D(p+1)-brane that
is localised in the transverse space labelled with
$\vec{\tR},\tY$. As in the case of a BPS D$p$-brane studied in
\cite{Kutasov:2004ct,Sen:2007cz} we will be interested in the
study of the dynamics of the mode $\tY$. In other words we firstly
solve the equations of motion for $\vec{\tR}$ and search the
solutions where $\vec{\tR}$ are constant. In fact, as follows from
the equations of motion for $\tilde{X}^m$ we obtain that they are
obeyed for
\begin{equation}
\tilde{X}^m=0 \ .
\end{equation}
In what follows we restrict to the study
of the action for the
$\tY$ and $T$ only. Note that for $\tR=0$
$h(\tR,\tY)$ is equal to
\begin{equation}
h(\tY,0)\equiv h(\tY)=\frac{k}{4\sin^2\frac{\tY}{2}} \ .
\end{equation}
Then the non-BPS D$(p+1)$-brane action takes the form
\begin{equation}\label{nonact3}
S=-\frac{\sqrt{2}\mT_{p+1}}{\tilde{g}}
\int d^{p+2}\xi
\frac{V(T)}{\sqrt{h(\tY)}}\sqrt{-\det(\eta_{\alpha\beta}
+h(\tY)\partial_{\alpha}\tY\partial_{\beta}\tY+
\partial_{\alpha}T\partial_{\beta}T)} \ ,
\end{equation}
where $V(T)$ is equal to
\begin{equation}
V(T)=\frac{1}{\cosh\frac{T}{\sqrt{2}}} \ .
\end{equation}
Now, following \cite{Kutasov:2004dj,Kutasov:2004ct}
we introduce
 tachyon field $\mT$ that is related to
$\tY$ through the relation
\begin{equation}
\frac{d\mT}{d\tY}=
\sqrt{h(\tY)}=\frac{\sqrt{k}}
{2\sin\frac{\tY}{2}} \  .
\end{equation}
This differential equation has the
 solution
\begin{equation}\label{costy}
\cos\frac{\tY}{2}=\frac{\sinh\frac{\mT}
{\sqrt{k}}}
{\cosh\frac{\mT}{\sqrt{k}}}\ ,
\end{equation}
where we have used the boundary condition
that for $\tY=\pi$ the  tachyon
field $\mT$ is equal to zero. Note
also that (\ref{costy}) implies
 that
for $\tY\rightarrow 0$ $\mT\rightarrow \infty$
while for $\tY\rightarrow 2\pi$ we obtain
that $\mT\rightarrow -\infty$. In other words
the new tachyon field belongs to the interval
$\mT\in (-\infty,\infty)$.
Then we obtain
\begin{equation}
h(\tY(\mT))=\frac{k}{4}\cosh^2
\frac{\mT}{\sqrt{k}} \
\end{equation}
and hence  the tachyon effective action
(\ref{nonact3})  can be written as
\begin{eqnarray}\label{mTT}
S&=&-\tau_{p+1}\int d^{p+2}\xi
\mathcal{V}(T,\mT)\sqrt{-\det(\eta_{\alpha\beta}
+\partial_{\alpha}\mT\partial_{\beta}\mT+
\partial_{\alpha}T\partial_{\beta}T)} \ ,
\nonumber \\
& &\mathcal{V}(\mT,T)=\frac{1}{ \cosh\frac{\mT}{\sqrt{k}}\cosh
\frac{T}{\sqrt{2}}}\equiv V(\mT,\frac{1}{k}) V(T,\frac{1}{2})
 \ , \nonumber \\
\end{eqnarray}
where
\begin{equation}\label{deftau}
\tau_{p+1}=\frac{2\sqrt{2}\mT_{p+1}}
{\sqrt{k}\tilde{g}} \ , \quad
V(f,x)\equiv \frac{1}{\cosh(f\sqrt{x})} \ .
\end{equation}
Now we come to the study of some
solutions of the  equations
of motion for $T$ and $\mT$
that arise from the action (\ref{mTT}).
Note also that these equations
has been also discussed  in \cite{Kluson:2004yk}.

Let us start with the solution that describes non-BPS
D$(p+1)$-brane wrapped transverse circle. As the first step we
determine  the equations of motion for $\mT,T$ from (\ref{mTT})
\begin{eqnarray}\label{eqmTT}
-\frac{\sinh\frac{\mT}{\sqrt{k}}}
{\sqrt{k}\cosh^2\frac{\mT}{\sqrt{k}}
\cosh\frac{T}{\sqrt{2}}}\sqrt{-\det
\mG}-
\partial_{\alpha}\left[
\frac{1}{\cosh\frac{\mT}{\sqrt{k}}
\cosh\frac{T}{\sqrt{2}}}\partial_\beta
\mT\mGi^{\beta\alpha}
\sqrt{-\det\mG}\right]=0\ , \nonumber \\
-\frac{\sinh\frac{T}{\sqrt{2}}}
{\sqrt{2}\cosh\frac{\mT}{\sqrt{k}}
\cosh^2\frac{T}{\sqrt{2}}}\sqrt{-\det
\mG}-
\partial_{\alpha}\left[
\frac{1
}{\cosh\frac{\mT}{\sqrt{k}}
\cosh\frac{T}{\sqrt{2}}}
\partial_\beta T\mGi^{\beta\alpha}
\sqrt{-\det\mG}\right]=0
\ . \nonumber \\
\end{eqnarray}
It is easy to see that for $k=2$ both the equations are
essentially the same. Let us now consider the situation when $\mT$ is
a function of one spatial variable on non-BPS D$(p+1)$-brane, say
$\xi^{p+1}\equiv x$ and construct the singular kink following the
analysis performed in \cite{Sen:2003tm}. First of all, the
equations of motion (\ref{eqmTT}) for $T=\mathrm{const}$ implies
that we have two solutions $T=T_{min}$ or $T=T_{max}$. We consider
the solution $T=T_{max}$ corresponding to unstable D$(p+1)$-brane.
Then the equation of motion for $\mT=\mT(x)$  takes the form
\begin{eqnarray}\label{eqmTT2}
-\frac{\sinh\frac{\mT}
{\sqrt{k}}
\sqrt{1+(\partial_x\mT)^2}}
{\sqrt{k}\cosh^2\frac{\mT}
{\sqrt{k}}}
-\partial_x\left(
\frac{1}{\cosh\frac{\mT}
{\sqrt{k}}}
\frac{\partial_x\mT}
{\sqrt{1+(\partial_x\mT)^2}}\right)
=0 \  \nonumber \\
\end{eqnarray}
that can be written as
\begin{equation}
\partial_x\left(
\frac{1}{\cosh\frac{\mT}
{\sqrt{k}}
\sqrt{1+(\partial_x\mT)^2}}\right)=0 \ .
\end{equation}
In other words  the expression in the bracket
above does not depend on $x$. Since
for a kink solution $\mT\rightarrow \pm \infty$
 as $x\rightarrow \pm \infty$ and $
V(\mT,k^{-1})\rightarrow 0$
 in this limit we obtain that
the expression in the bracket vanishes for
$x\rightarrow \infty$ and from its independence
on $x$ it implies that it vanishes everywhere.
This in turn implies that we should have
\begin{equation}
\mT=\pm \infty \ \mathrm{or}
\ \partial_x\mT=\infty \
\mathrm{(or \ both)} \   \mathrm{for \ all} \ x \ .
\end{equation}
Clearly this solution looks singular.
Again, following \cite{Sen:2003tm}
we  consider the field configuration
\begin{equation}\label{fd}
\mT(x)=f(ax), \quad
f(u)=-f(-u) , \quad  f'(u)<0 \quad  \forall x, \quad
f(\pm \infty)=\mp \infty
\end{equation}
that in the limit $a\rightarrow \infty$
is singular.
For this solution however we get that
\begin{equation}
\frac{V(f(ax),k^{-1})}{
\sqrt{1+(\partial_x \mT)^2}}
\approx
\frac{V(f(ax),k^{-1})}{
a|f'(ax)|}
\end{equation}
that vanishes everywhere at
the limit
$a\rightarrow \infty$ since the numerator
vanishes (except at $x=0$) and the
denominator blows up everywhere.
For next purposes it will be also
useful to  calculate the
stress energy tensor
from the action (\ref{mTT}).
Following the analysis outlined
in previous section we obtain
\begin{equation}\label{TconR}
T_{\alpha\beta}=-\tau_{p+1}
\mV(T,\mT)\eta_{\alpha\gamma}
\eta_{\beta\delta}\mGi^{\delta\gamma}
\sqrt{-\det \mG} \ .
\end{equation}
Then we obtain following components of the
stress energy tensor
\begin{eqnarray}
T_{ij}(x)&=&-\eta_{ij}
\tau_{p+1}V(\mT,k^{-1}
)\sqrt{1+(\partial_x\mT)^2}=\nonumber \\
&=&
-\eta_{ij}\tau_{p+1}
V(f(ax),k^{-1})
\sqrt{1+ a^2f'^2(ax)}\approx
\nonumber \\
&\approx&-\eta_{ij}\tau_{p+1}
V(f,k^{-1})
|af'(ax)| \, \quad i,j=0,\dots,p \nonumber \\
\end{eqnarray}
 in the limit $a\rightarrow \infty$.
Then the integrated $T_{ij}$
 associated
with the codimension one solution are equal
to
\begin{eqnarray}
T^{kink}_{ij}&=&
\int dx T_{ij}=
-\eta_{ij}\tau_{p+1}
 \int dx
V\left(f,\frac{1}{k}\right)
|af'(ax)|=\nonumber \\
&=&
-\eta_{ij}\tau_{p+1}
 \int dm V\left(
m,\frac{1}{k}\right) \ ,
\nonumber \\
\end{eqnarray}
where $m=f(ax)$.
Thus $T^{kink}_{ij}$
depend
on $V$ and not on the form of $f(u)$.
Note that for $V(f,k^{-1})$ defined
in (\ref{deftau}) we obtain
\begin{equation}\label{in}
\int dm\frac{1}
{\cosh \frac{m}{
\sqrt{k}}}=\sqrt{k}\frac{(2\pi)}{2} \ .
\end{equation}
 In fact,
using (\ref{in})
we obtain that the  tension of the resulting
kink is equal to
\begin{equation}
\tau_{p+1}\sqrt{k}\frac{(2\pi)}
{2}=2\pi \sqrt{2}\mT_{p+1}\tilde{g}^{-1}=
\sqrt{2}\tilde{g}^{-1}\mT_p
 \ .
\end{equation}
Let us give the geometrical meaning of this solution. Since $\mT$
is directly related to the coordinate  $\ty$ that parameterises
the position of a non-BPS D$(p+1)$-brane on the transverse circle,
the singular kink solution corresponds to non-BPS D$(p+1)$-brane
that sits on the top of the five branes for all $x<0$,  at $x=0$
wraps the $\ty$ circle and then back to the five branes at
$\ty=2\pi $ where it stays for all $x>0$. In other words this
solution describes non-BPS D$p$-brane wrapping around the
transverse circle that was recently discussed in
\cite{Sen:2007cz}.

Let us now consider following ansatz
\begin{equation}
\mT=\mathrm{const}, \quad T(x)=f(ax)
 \ .
\end{equation}
Firstly, the solutions of the equation of motion for constant
$\mT$ is either $\mT=\pm\infty$ or $\mT=0$. The case $\mT=\pm
\infty$ corresponds to non-BPS D$(p+1)$-brane sitting on
world-volume of NS5-branes. On the other hand the case $\mT=0$
corresponds to non-BPS D$(p+1)$-brane sitting at the point $
\tY=\pi$. Let us consider this situation where now
$V\left(\mT,\frac{1}{k}\right)=1$. Then the invariance of the
world-sheet energy tensor implies that
\begin{eqnarray}
\partial_x T_{xx}&=&0 \ ,
\quad T_{xx}=-\tau_{p+1}\mV(\mT,T)
\mGi^{xx}\sqrt{-\det\mG}=\nonumber \\
&=&-\tau_{p+1}
V(f(ax),\frac{1}{2})
\frac{1}{\sqrt{1+a^2f'^2(ax)}}\approx
\nonumber \\
&\approx&
-\tau_{p+1}
\frac{V(f(ax),\frac{1}{2})}
{a|f'(ax)|}
 \ .
\nonumber \\
\end{eqnarray}
Again standard arguments imply that the
$T_{xx}$ has
to vanish.
On the other hand the remaining components
of $T_{\alpha\beta}$ are equal
to
\begin{eqnarray}\label{Tija}
T_{ij}&=&-\eta_{ij}\tau_{p+1}
V(f(ax),\frac{1}{2})
\sqrt{1+a^2f'^2(ax)
}\approx
\nonumber \\
&\approx &-\eta_{ij}\tau_{p+1}
V(f(ax),\frac{1}{2})
a|f'(ax)| \ .
\nonumber \\
\end{eqnarray}
We  define
 the stress energy tensor
of the kink as the integration
of (\ref{Tija}) over $x$
\begin{eqnarray}
T_{ij}^{kink}&=&\int dx  T_{ij}(x)=
-\eta_{ij}\tau_{p+1}
\int dx
V(f(ax),\frac{1}{2})a|f'(ax)|=
\nonumber \\
&=&-\eta_{ij}\tau_{p+1}
 \int dm V(m,\frac{1}{2})=
-\eta_{ij}\frac{2\mT_p}{
\sqrt{k}}\tilde{g}^{-1}
\nonumber \\
\end{eqnarray}
using
\begin{equation}
\sqrt{2}\mT_{p+1}\int dm V(m,\frac{1}{2})=
\mT_p \ .
\end{equation}
The geometrical meaning of this solution is clear. It is the
$G$-type D$p$-brane sitting at the point $\ty=\pi$.

In summary, we have shown that non-BPS D$(p+1)$-brane that is
stretched along the world-volume directions of the NS5-brane
contains as its solution the non-BPS D$(p+1)$-brane that wraps the
transverse circle and also contains solution corresponding to
$G$-type D-brane that sits at the point $\ty=\pi$. Note also that
for $k=2$ the fields $\mT$ and $T$ have completely the same
dynamics. Since these two D-branes can be interpreted as solutions
of one world-volume theory they are-from the point of view of this
theory-indistinguishable. In other words, for $k=2$, the unstable
D$(p+1)$-brane wrapping transverse circle and the $G$-type
D$p$-brane sitting at the point $\ty=\pi$ are actually the same
object. We would like to mention that this fact can serve as a
support of the Sen's recent conjecture.
\section{Conclusion and Discussion}
In this paper we have studied the non-BPS branes from the view
point of DBI analysis in NS5-brane background. We have presented
two complimentary description of the non-BPS D-brane in NS 5-brane
background depending on whether the brane wraps the transverse
circle of extended along the NS5-brane worldvolume directions. We
have shown that when the non-BPS D-brane wraps the transverse
circle along the NS 5-brane, there exist tachyon solutions that
reproduce the unstable D-brane solutions of \cite{Sen:2007cz}.
We have also argued that the physical interpretation is
not completely clear due to the presence of the infinite factor
corresponding to the value of the tachyon at its minimum.

The second description has the non-BPS branes stretched along the
longitudinal directions on the NS 5-branes. The main advantage of
this approach with respect to the previous one is that now the
mode that parametrises the position of non-BPS D$(p+1)$-brane
along transverse circle can be mapped to geometric tachyon field
$\mT$ that has similar properties as the ordinary one $T$.
 Then it is easy to construct the solution given in
\cite{Sen:2003tm}. Explicitly, we  have shown that this theory
contains a non-BPS D$(p+1)$-brane that wraps the transverse circle
$(y)$, along with a solution which can be seen as a D$p$-brane
with a geometrical instability (the $G$-type D-brane) localised at
$y= \pi R$. This is a hint of the unified description of the
$G$-type D-brane and the $U$-type D-brane arising out of the
worldvolume theory on the non-BPS D$(p+1)$-brane localised on
$y$-circle. Then we have studied the situation in the small radius
limit. We show that the correspondence between the $G$-type and
$U$-type branes is also obeyed. We have further shown the
equivalence of the geometric tachyon field and the usual open
string tachyon field and have shown that for the value $k=2$ they
essentially have the same dynamics. For this value the unstable
D-brane which wraps the transverse circle and the $G$-type D-brane
which  sitting at $\tilde y = \pi$ are actually the same object.

We would also like to suggest an extension of the present work.
Firstly, it would be interesting to study the properties of the
resulting solutions from the Hamiltonian point of view. In fact,
it is well known from the Hamiltonian analysis of the unstable
Dp-branes that many new interesting phenomena occur, for example
the emergence of the closed strings
\cite{Yee:2004ec,Kwon:2003qn,Sen:2003bc,
Sen:2002qa,Gibbons:2000hf,Sen:2000kd,Yi:1999hd,Lindstrom:2001qa}.
It would be extremely interesting to see related phenomena in case
of geometric tachyon and find their physical interpretation. We
hope to return to this problem in future.

Before ending the paper, we would like to mention few things about
the dual ALF theory as discussed in \cite{Sen:2007cz}. The non-BPS
D$(p+1)$-brane of the original theory along $x^0,\dots,x^p,y$
placed at $\vec{r}$ are easy to describe in the dual theory. This
goes over to a non-BPS D$p$-brane along $x^0,\dots,x^p$ in the
dual system placed at fixed values of $\vec{r}$ and $\psi$ in the
ALF space, with the location along $\psi$ determined by the Wilson
line along $y$ of the original system. On the other hand let us
consider non-BPS D$(p+1)$-brane in the original theory that is
sitting at particular point in $y$. In the dual theory it wraps
$\psi$ circle. However we immediately come to the puzzle that has
the same origin as in case of BPS D$p$-brane. Namely, G-type
unstable D-branes that in the original description correspond by
placing BPS D$p$-brane along $x^0,\dots,x^p$ at $(\vec{r},y=0)$ or
$(\vec{r}=0,y=\pi R)$. Since T-duality acting on a D-brane
localised at a point on a circle maps it to a D-brane wrapped
along the dual circle we expect that the $T$-dual description of
the $G$-type brane is BPS D$(p+1)$-brane along $x^0,\dots, x^p$
and $\psi$ and placed at $\vec{r}=0$. The coordinate $y$ of the
original D$p$-brane corresponds to the Wilson line along $\psi$ on
the dual D$(p+1)$-brane. However at the level of supergravity
approximation we do not see a potential for the Wilson line.
However they are expected to be induced by the world-sheet
instanton corrections
\cite{Tong:2002rq,Harvey:2005ab,Okuyama:2005gx}.
It would be again extremely interesting whether we can
find field redefinition that can maps this potential
and the Wilson line variable to the tachyon-like form.

\vskip .5in \noindent {\bf Acknowledgements:}
 The work of JK was supported in part by the Czech
Ministry of Education under Contract No. MSM 0021622409, by INFN,
by the MIUR-COFIN contract 2003-023852, by the EU contracts
MRTN-CT-2004-503369 and MRTN-CT-2004-512194, by the INTAS contract
03-516346 and by the NATO grant PST.CLG.978785.


\end{document}